\documentclass[a4paper,11pt]{amsart}

\usepackage{verbatim} 
\usepackage[utf8]{inputenc}
\usepackage{mathbbol,mathrsfs,amsfonts,amssymb,amsthm,stmaryrd}
\usepackage{amsmath}
\usepackage[all]{xypic}
\usepackage{mathpazo}
\usepackage[mathcal]{euscript}
\usepackage{graphicx, floatflt}
\usepackage{multicol}
\usepackage{nicefrac}
\usepackage{caption}

\usepackage{booktabs}

\usepackage{geometry}
\usepackage[colorlinks,hyperindex]{hyperref}
\hypersetup{linkcolor=blue, citecolor=blue}

\newcommand{\abs}{{\rm abs}}

\theoremstyle{theorem}

\begin{document}

\title[Energy Estimates of PFC and the WTC Catastrophe]{Energy Estimates of Progressive Floor Collapses \\
and the World Trade Center Catastrophe}

\author[Ansgar Schneider]{Ansgar Schneider}

\begin{abstract}
The Simple Collapse Model of Bažant and Zhou is evaluated
for a progressive floor collapse of a tall building. 
A sequence of energy estimates indexed by the collapsing floors is derived.
Each of the estimates gives a sufficient condition to arrest the collapse 
at a given floor.

The first estimate of this sequence has been stated by 
Bažant and Zhou and has been repeatedly cited later on.
However, this estimate is not 
optimal in the sense that the following estimates 
give a weaker condition to arrest the collapse.
\end{abstract} 
\maketitle


\tableofcontents

\sloppy
\section{Introduction}
\noindent
On the 11th of September 2001 both the North and the South Tower of the World Trade Center in New York City, USA,
collapsed after both of them were struck by an aircraft. 
In this short note we shall solely  focus on the collapse of the North Tower.
The building had 110 storeys and a 
roof height of  417\,m.
The fuselage of one of the two aircrafts impacted the North Tower at the height of the 96th floor. 
The Tower collapsed 102 minutes later \cite{NIST1}.

Two days after the attacks Bažant and Zhou submitted a paper 
to explain why the collapse might have occurred after the (total) failure of just one storey \cite{BaZh02}.
A gravity-driven progressive floor collapse is proposed as an explaination. This analysis is based on their 
Simple Collapse Model where the falling top section of the building descents through one storey in free fall and 
impacts the floor below inelastically.   
Their main result was a sufficient condition for the energy absorption capacity 
of the buckling columns in the floor below to arrest the fall. 
Based on the numerical values they had at the time
they found this condition was violated by almost one order of magnitude (a factor 8.4) for the 
North Tower and concluded that 
the collapse became inevitable once the  failure of only 
one storey appeared.

This result has been repeatedly stated in the following works 
of Bažant and Verdure \cite{BaVe07} and of Bažant, Le, Greening and Benson \cite{BBGL08}
and also in \cite[p.~323]{NIST1-6}
without being corrected. Indeed, two things need an explicit clarification:
\begin{enumerate}
\item The numerical values of the quantities that are used for evaluating the estimate numerically. 
In particular these are
the mass of the falling top section, the maximal possible energy dissipation of the buckling columns,
and the resistance through the first failing storey.
\item The  estimate given in \cite{BaZh02} does not give the minimal value of energy dissipation 
that would be sufficient to arrest the collapse. The minimal value is lower.
\end{enumerate}
The numerical values have been corrected by others, and we shall recapitulate 
some of the discussion below.
The main objective of this short note is to explain the second one of the two mentioned points.

\section{Simple Collapse Model}
\noindent
Let us carefully review the Simple Collapse Model proposed in \cite{BaZh02}
and clarify the critical condition for a global collapse.

\subsection{Model Assumptions} 
\begin{figure}[b]
	\includegraphics[scale=0.45]{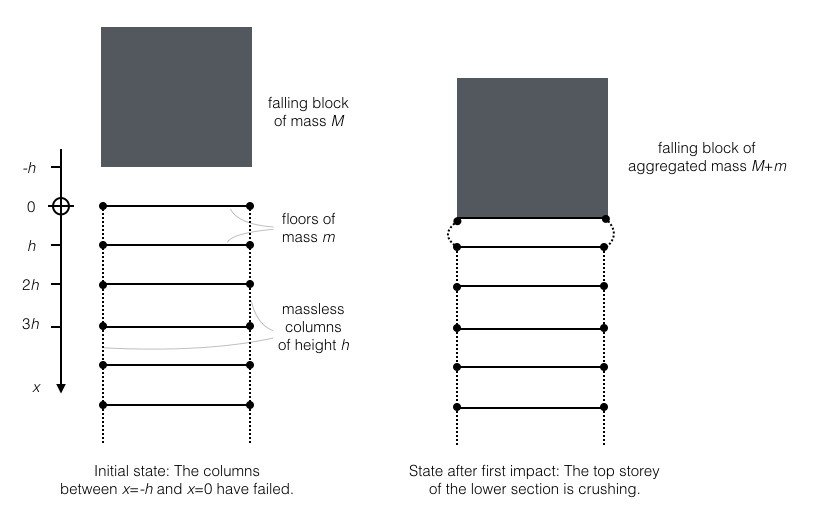}
	\caption{Simple Collapse Model}
	\label{Pic1}
\end{figure}

\noindent
We consider a collapse sequence whose beginning is illustrated in Figure \ref{Pic1}. 
The top section (of mass $M$) of a tall building  crushes on the lower part due to column failure in a certain storey.
Let us first assume a worst case scenario in which the top section of the building 
encounters no resistance during the fall through the storey that initially failed.
We assume the following sequence of events and properties:
\begin{itemize}
\item All storeys have the same hight $h$. All floors have the same mass $m$, but no vertical extension. 
(Throughout this short note we use the word `floor' for the massive bottom part of each storey.)

\item All columns of the building are massless and once they are broken they don't contribute to the collapse any more.

\item The columns in each storey have the same strength. So the amount of energy $E_\abs$ that is needed for bending/crushing the columns in one storey is the same for all storeys.

\item The crushing front propagates from top to bottom in the following sense. 
The impact on the floor at position $x=0$  in Figure  \ref{Pic1} is assumed to be inelastic. 
After impact the top section and that floor move downwards as a 
single block of aggregated mass $M+m$. If the energy dissipation of the buckling columns 
is too low to arrest the fall,
the falling section impacts the floor at position $x=h$. This impact is again inelastic and 
results in a an aggregated falling mass of $M+2m$.
Then the propagation of the collapse might continue as described.
\end{itemize}

\subsection{Remark on the Model Assumptions}
The assumption that the top section is not destroyed during the collapse is based on the 
argument given in \cite[Appendix]{BBGL08} where a two-sided front propagation 
is computed in a similar set-up. 
Therein the upward-directed front terminates right after collapse initiation.  

\subsection{Critical Energy Absorption Capacity}
\label{SubSecCapacity1}
Let us denote by $v_0$ the velocity of the top block of mass 
$M$ right before impact on the first floor at position $x=0$ in Figure \ref{Pic1}.
By conservation of energy it satisfies 
\begin{eqnarray}
\label{E-Satz1}
\frac{1}{2}Mv_0^2= Mg h,
\end{eqnarray}
 where $g$ is the acceleration of gravity and $h$ is the height of one storey. 
After impact on the first floor of mass $m$ the top section and the first floor move downwards as a 
single block of aggregated mass $M+m$. The velocity right after impact can be computed 
by conservation of momentum and is given by
\begin{eqnarray}
u_0 = \frac{M}{M+m} v_0 = \frac{a_0}{a_1} v_0,
\end{eqnarray}
where here and in the following we use the convention $a_k:= 1+ k\frac{m}{M}$, for any number $k=0,1,2,3, \dots$.
The motion of the block with mass $M+m=Ma_1$ will stop before reaching position $x=h$
if bending and crushing of the columns between $x=0$ and $x=h$ absorbs more energy as given 
by the kinetic energy of the mass $Ma_1$ and its loss of potential energy.
In other words, if the energy absorption capacity $E_\abs$  of the columns of one storey 
satisfies the inequality
\begin{eqnarray}
\label{Cond1}
E_\abs \geq \frac{1}{2}(Ma_1) u_0^2 + (Ma_1) g h = Mgh \frac{1+a_1^2}{a_1},
\end{eqnarray}
then the fall will arrest before the position $x=h$ is reached. 
However, if inequality (\ref{Cond1}) is violated, then the collapse  might progress. 
(This is the case if the maximal resistance force of the columns between $x=h$ and $x=2h$ is reached during the impact. 
The columns will then deform plastically and will eventually buckle if the load is too big.) 
If the collapse progresses, then the
velocity $v_1$ of the aggregated mass $Ma_1$ right before hitting the next floor can 
again be computed by energy conservation which takes the form
\begin{eqnarray}
\label{E-Satz2}
\frac{1}{2}M a_1v_1^2 =  Mgh \frac{1+a_1^2}{a_1} -  E_\abs.
\end{eqnarray}
Then again the impact on the floor at position $x=h$ will be assumed to be inelastic and the falling block will have mass $M+2m=Ma_2$.
Conservation of momentum implies that the velocity $u_1$ right after the impact is given by
\begin{eqnarray}
u_1= \frac{M+m}{M+2m}v_1= \frac{a_1}{a_2} v_1.
\end{eqnarray}
Similar as above we  conclude that the motion of the newly aggregated block with mass $Ma_2$ will arrest before position $x=2h$
if the inequality
\begin{eqnarray}
E_\abs &\geq& \frac{1}{2}(Ma_2) u_1^2 + (Ma_2) g h\\
&=&\frac{a_1}{a_2}\cdot\frac{1}{2}M a_1 v_1^2 + M g h a_2\nonumber
\end{eqnarray}
holds.  Inserting (\ref{E-Satz2}) and solving the resulting inequality for $E_\abs$ gives
\begin{eqnarray}
\label{Cond2}
E_\abs \geq  Mgh \frac{1+a_1^2+a_2^2}{a_1+a_2}.
\end{eqnarray}
Note that if the fraction $m/M$ is small then the fraction $(1+a_1^2)/a_2$ is approximately 2, whereas the 
fraction $(1+a_1^2+a_2^2)/(a_1+a_2)$ is approximately $3/2$, so condition (\ref{Cond2}) is indeed weaker than condition (\ref{Cond1}).

It is left to the reader to repeat the arguments from above in an easy induction by $n$
to conclude that the inequality
\begin{eqnarray}
\label{Condn}
E_\abs \geq  Mgh \frac{1+\sum_{k=1}^n a_k^2}{\sum_{k=1}^n a_k}.
\end{eqnarray}
is a sufficient condition to arrest the fall before position $x=nh$. 
It is clearly not a necessary condition, because the actual column 
forces are not reflected in this consideration. 

The first one  of these estimates ($n=1$) has been  
mentioned in  \cite{BaZh02,BaVe07,BBGL08}, where on 
its basis it is argued that a gravity-driven progressive floor collapse of the Twin Towers 
was inevitable (cp.~Section~\ref{Conclusion}).

\subsection{Evaluation of Critical Capacity}
Using the summation formulas for the numbers from $1$ to $n$ and for the
square numbers from $1$ to $n^2$ the fraction of the right hand side can be made explicit:
\begin{eqnarray}
\frac{1+\sum_{k=1}^n a_k^2}{\sum_{k=1}^n a_k} &=& 
\underbrace{\frac{1+n+ n(n+1)\frac{m}{M} + \frac{1}{6}n(n+1)(2n+1)\left(\frac{m}{M}\right)^2}{n+\frac{1}{2}n(n+1)\frac{m}{M}}}
\\\nonumber
&&\qquad\qquad\qquad\qquad\qquad=:F\left(n, \frac{m}{M}\right).
\end{eqnarray}
The collapse of the North Tower of the World Trade Center originated at the 
98th floor \cite[p.\,151]{NIST1}. The fuselage of the aircraft crushed into the 96th floor. The building had 110 storeys, so there were 12 or 13 storeys  (or maybe up to 15, see the discussion about Figure 19 in \cite[Section 2.3]{Schn17}) plus the roof with its antenna above the failing floor. We therefore consider $0.077\simeq 1/13$ (or $0.063\simeq 1/16$) to be a reasonable value of the fraction $m/M$. 
For $n=1,2,\dots 10$ the corresponding values of  $F(n,\nicefrac{m}{M})$ are given in Table~\ref{FirstResults}.
\begin{center}
  \begin{tabular}{   c  c   c  c  c  c  c  c  c  c  c  }
  \\
\toprule
    $n$ & 1 & 2 & 3 & 4 & 5 & 6 & 7 & 8 & 9  &10 \\ \midrule
    $ F(n, 0.077)$ & 2.01 & 1.56 & 1.45 & 1.41 & 1.40 & 1.41 & 1.44 & 1.46 & 1.49 & 1.53\\ 
    $ F(n, 0.063)$ & 2.00 & 1.55 & 1.42 & 1.38 & 1.36 & 1.37 & 1.38 & 1.40 & 1.42 & 1.45\\
\bottomrule
\\
  \end{tabular} 
\captionof{table}{Some values of $F$.}
\label{FirstResults}
\end{center}
The minimal value of $F(n, 0.077)$ is 1.40 for $n=5$. 
In the Simple Collapse Model this means: If the top section has 12 to 13 storeys, then
a floor-wise energy absorption capacity that is 40\% higher 
than the initial kinetic energy gained during the fall through the initial failing storey
is sufficient to arrest the fall (before position $x=5h$).

\subsection{Model Refinement}
It is reasonable to assume the initally failing storey 
does absorb some energy. So the initial kinetic energy 
might be given as
\begin{eqnarray}
\label{E-SatzAlpha}
\frac{1}{2}Mv_0^2=\alpha Mgh,
\end{eqnarray}
where $\alpha\in [0,1]$. It is demanded in \cite{BaLe11} that $\alpha$
should be at least $0.794$. 
However, actual measurements of the acceleration of the roofline 
of the North Tower show that  a lower value actually appeared \cite{MaSz09, Chan10}. 
During the first 1.2 seconds, i.\,e.~during the 
fall through the first storey the acceleration was $0.52\,g$ \cite[Figure~1]{JSS13}.
Anticipating this empirical datum, the computation done in Section~\ref{SubSecCapacity1} can be redone exactly as before
but with (\ref{E-Satz1}) replaced by (\ref{E-SatzAlpha}). The resulting  energy estimate then 
becomes
\begin{eqnarray}
\label{CondnAlpha}
E_\abs \geq  Mgh F_\alpha\left(n,\frac{m}{M}\right),
\end{eqnarray}
where
\begin{eqnarray}
F_\alpha\left(n,\frac{m}{M}\right):=\frac{\alpha+\sum_{k=1}^n a_k^2}{\sum_{k=1}^n a_k}.
\end{eqnarray}
For $\alpha=0.52$ and $\alpha=0.79$ we find the first ten values of $F_{\alpha}(n,0.077)$ in Table~\ref{BetterRes}.
\begin{center}
  \begin{tabular}{   c  c   c  c  c  c  c  c  c  c  c  }
\\
\toprule
    $n$ & 1 & 2 & 3 & 4 & 5 & 6 & 7 & 8 & 9  &10 \\ \midrule
    $ F_{0.79}(n, 0.077)$ & 1.81 & 1.47 & 1.39 & 1.36 & 1.37 & 1.39 & 1.41 & 1.44 & 1.48 & 1.51\\
    $ F_{0.52}(n, 0.077)$ & 1.56 & 1.35 & 1.31 & 1.31 & 1.33 & 1.35 & 1.38 & 1.42 & 1.46 & 1.49\\
\bottomrule
\\
  \end{tabular} 
\captionof{table}{Some values of $F_{0.52}$ and $F_{0.79}$.}
\label{BetterRes}
\end{center}

In \cite{BaVe07} the compaction parameter $\lambda$ is introduced to 
describe the size of the crushed building part relative to its original size.
A value of $\lambda=0.18$ is used in \cite{BaVe07} a value of $\lambda=0.15$ 
is used in \cite{Schn17}.
We can easily include this parameter in the Simple Collapse Model by giving 
an extension of $\lambda h $ to each floor. In other words, if one storey is crushed
the falling section of the building has descended by a height of  $(1-\lambda)h$. 
(The impact of two floors is still assumed to be inelastic.)
Therefore the energy estimates we obtain are just the same as before 
but rescaled by the factor $1-\lambda$.

Including $\alpha$ and $\lambda$  a sufficient condition 
to arrest the fall is given by
\begin{eqnarray}
\label{CondnAlphaLambda}
E_\abs \geq  Mgh (1-\lambda) F_\alpha\left(n,\frac{m}{M}\right).
\end{eqnarray}

\section{Discussion of Results}

\subsection{Numerical Evaluation}
In \cite{BaZh02} a mass of $M=58\cdot 10^6\,\rm kg$
is stated without reference. Two related quantities 
that are also stated without reference in \cite{BaVe07,BBGL08}
are the total mass of the North Tower, 500,000\,t, and
the height of the falling top section, 80\,m.

It has  already been pointed out in \cite{JSS13} that
$M=58\cdot 10^6\,\rm kg$ is too big and a
value of $M=33\cdot 10^6\,\rm kg$ has been used
based on \cite[Table~4-7]{NIST1-6D}.

According to \cite[p.\,151]{NIST1} and \cite[Sec.~2.3]{Schn17}
the falling top section had a height of 46 to 55\,m.
In \cite{Schn17} a mass density of $0.6\cdot 10^6\,\nicefrac{\rm kg}{\!\rm m}$
is used for the top 30 storeys. This value is 
based on a total mass of 288,000\,t of the tower, 
which is the value that has been estimated in \cite{Uri07}.
A height of $50\,\rm m$ (approx.~13 storeys with a height of $h=3.8\,\rm m$) for the top section then gives 
$M=30\cdot 10^6\,\rm kg$.

So if we evaluate (\ref{CondnAlphaLambda}) with 
the the bigger value of $M$ we find that  
\begin{eqnarray}
\label{Important}
E_\abs&\ge& 33\cdot 10^6 \rm kg\cdot 9.8\,\nicefrac{\rm m}{\sec^2}\cdot 3.8\,\rm m\cdot (1-0.15) \cdot 1.31\\\nonumber
&=& 1370\, \rm MJ
\end{eqnarray}
is a sufficient condition to arrest the fall.
Note that for $\alpha=1$ the result is just 1460\,MJ, 
which is less then 7\% bigger than the value of (\ref{Important}).
 
\subsection{Empirical Values of Energy Dissipation}
In \cite{BaZh02} a maximal energy dissipation 
per storey of 500\,MJ is stated. This value is based on 
computations for a three-hinge buckling model for the 
crushing columns. 
However, Korol and Sivakumaran have conducted empirical 
studies of buckling columns and found that this value 
must be corrected by a factor 3 to 4 \cite{KS14}. 
This means a range from 1500\,MJ to 2000\,MJ
should be regarded as a maximal possible value
of energy dissipation per storey.
As this range  exceeds the value of (\ref{Important}), the conclusion 
that the collapse was inevitable is wrong if it is based 
on the Simple Collapse Model. 
For the higher value of 2000\,MJ 
this is still correct if a structural damage of the columns 
of over 25\% is assumed. (This statement is also correct for $\alpha=1$.)

In \cite{BaLe16} it is demanded that the empirical values of \cite{KS14} 
should be rescaled by a factor of 2/3 which would give 
a range of 1000\,MJ to 1300\,MJ.
If this range is correct, no definite statement can be made
within the uncertainty of the other relevant quantities.
So it is certainly false to claim that
a priori the design of the columns was too weak to arrest the fall. 

It should be emphasised at this point that so far we are discussing the 
maximal possible amount of energy dissipation per storey.
This value appears if all columns buckle according to the three-hinge model.
However, this value does not match the observed values during the actual collapse.
During the first 4.6 seconds of the collapse the energy dissipation per storey was below 250\,MJ \cite[Section~2.4, Figure~4]{Schn17}.
Moreover, between 4.6 and 7.7 seconds after collapse initiation   a value that corresponds 
to (at least) 2000\,MJ of energy dissipation per storey 
must have occurred if the collapse was gravity-driven \cite[Section~2.7, Figure~8]{Schn17}.
The observation of this high value of energy dissipation implies  
that \emph{if} the maximal possible of energy dissipation was only in a range 
of 1000\,MJ to 1300\,MJ, 
then the collapse was not gravity-driven.\footnote{
Here we use the term ``gravity-driven collapse" as a shorthand for a collapse 
that is described by the 
Crush-Down equation, which essentially is a continuous (non-discrete)
version of the presented Simple Collapse Model. 
It might be regarded as the limit $h\to 0$ with constant total mass and total height of the building.
The Crush-Down equation has been proposed and modified in \cite{BaVe07,BBGL08}.
A discussion and correction of some of the terms is given in \cite{Schn17} and \cite{Schn17b}. 
}

\subsection{Conclusion}
\label{Conclusion}
In \cite{BaZh02} a non-optimal estimate (n=1)
was evaluated with incorrect numerical values. This led the authors to the 
conclusion that the energy absorption capacity of the 
buckling columns of the North Tower was too low  to arrest the initiated 
collapse of the building by a factor of 8.4. 
Based on this erroneous result it is stated in \cite{BBGL08}: 
\begin{itemize}
\item[]
\textit{“Merely to be convinced of the inevitability of [a] gravity[-]driven progressive collapse, further analysis is, for a structural engineer, superfluous. Further analysis is nevertheless needed to dispel false myths, and to acquire [a] full understanding that would allow assessing the danger of [a] progressive collapse in other situations."
}
\end{itemize}
We hope that the structural engineer to whom this statement refers is not involved in any real-life construction,
for he might show a lack of critical thought in other situations, too. Yet much more than we are concerned about that  
one structural engineer, we are concerned about the dogmatic attitude  
that is manifest in these formulations.

Nonetheless we very much agree with the second one of these three propositions, to which 
we devote this work in accordance with its principal theme.

\end{document}